# OPTIMIZATION OF TIME DATA CODIFICATION AND TRANSMISSION SCHEMES: APPLICATION TO GAIA

J. Portell[1,2], E. García-Berro[1,2] and X. Luri[1,3]

[1] Institut d'Estudis Espacials de Catalunya, c/ Gran Capità, 2-4 (Edif. Nexus). 08034 Barcelona, Spain
[2] Dept. Física Aplicada, Universitat Politècnica de Catalunya, Escola Politècnica Superior de Castelldefels, Av. del Canal Olímpic, s/n. 08860 Castelldefels, Spain
[3] Dept. Astronomia i Meteorologia, Universitat de Barcelona, Facultat de Físiques, c/ Martí i Franquès, s/n. 08028 Barcelona, Spain
E-mail addresses: portell@ieec.fcr.es, garcia@fa.upc.es, xluri@am.ub.es

**Abstract.** Gaia is an ambitious space observatory devoted to obtain the largest and most precise astrometric catalogue of astronomical objects from our Galaxy and beyond. On-board processing and transmission of the huge amount of data generated by the instruments is one of its several technological challenges. The measurement time tags are critical for the scientific results of the mission, so they must be measured and transmitted with the highest precision – leading to an important telemetry channel occupation. In this paper we present the optimisation of time data, which has resulted in a useful software tool. We also present how time data is adapted to the Packet Telemetry standard. The several communication layers are illustrated and a method for coding and transmitting the relevant data is described as well. Although our work is focused on Gaia, the timing scheme and the corresponding tools can be applied to any other instrument or mission with similar operational principles.

## 1. Introduction

Gaia is one of the most ambitious space astrometry mission currently envisaged, adopted within the scientific programme of the European Space Agency (ESA) in October 2000, to be launched around 2010-2012. It aims to measure several features of an extremely large number of stars and astronomical objects with unprecedented accuracy, such as their positions and proper motions, photometry, and radial velocities. As a result, a three-dimensional map of more than 1 billion stars of our Galaxy will be obtained, as well as solar system objects and extragalactic sources. The precision of the angular measurements will be about 10 microarseconds (µas) at a moderate brightness ($15^{th}$ magnitude). In order to achieve this, Gaia will use two telescopes combined onto a single focal plane, named Astro, composed of 180 state-of-the-art Charge Coupled Devices (CCDs). The satellite will continuously scan the sky, allowing for about 85 observations of each star during the 5 years of duration of the mission. Each observation will be composed of 16 measurements, performed by 11 CCDs of an astrometric field (AF) and 5 CCDs of a broad band photometer (BBP). These are some of the specifications given by the current design, but there may be small changes in the future. Full sky coverage will be possible because of the spin of the satellite around its own axis, which itself precesses at a fixed angle of 50° with respect to the Sun. A third telescope projects the sources onto another instrument, named Spectro, composed of two CCD focal planes that will obtain spectra and medium band photometry, besides the angular (astrometric) measurements. This will lead to the most complete and accurate map of the stars of our Galaxy.

A continuous scanning of the sky using CCDs implies a special imaging operation, different than the typical shutter-based one. The best option for this case is the so-called time delayed integration (TDI), based on a continuous charge shift from one pixel row to the next, synchronized with the satellite spin motion. It leads to an intrinsically quantised operation of the CCDs, which eases the operation of time tagging their measurements. Long exposure times can be achieved in this way with minimum distortion and blur, being the total transit time of a stellar source about 57 seconds in the astrometric focal plane – during which 11 high-resolution measurements of 3.3 seconds each will be performed.

Measuring more than 1 billion objects several times with the highest resolution implies a technological challenge, not only for the predicted 20 TB compressed raw data transmitted to ground, but also – and specially – for the on board data handling. The development of a selective sampling method is mandatory for reducing the on-board video data rate from about 7 Gbps to a few hundredths of Mbps. It can be achieved by detecting and selecting the most relevant sources to measure, reading only *windows* (sets of pixels) around them. Despite of this, the on-board data handling system must reduce yet much more this amount of data by using optimised codification and transmission schemes – besides the lossless data compression system. It is the only way to download all of the data using the 1.7 Mbps available downlink. This average capacity is the result of an instantaneous 4 Mbps data link, combined with the 10 hours daily visibility of Gaia from the ground station – since the satellite will orbit around the L2 lagrangian point, 1.5 million kilometres from the Earth opposite to the Sun. The Cebreros ground station in Spain is, presumably, the only radio station to be used for receiving the data, so a permanent link between Gaia and the ground station will not be available. This implies that all the data acquired by the astronomical instruments must be stored on a mass storage system, waiting for being transmitted to ground during the next contact. These are the relevant features of Gaia for our work. More details on the mission can be found at *http://sci.esa.int/gaia/*.

In this article we list the main timing requirements of the mission and its instruments, which must be fulfilled by the time data codification method developed in our work. In doing so, the latest design of Gaia will be taken into account, as well as different sampling options depending on the brightness or kinds of sources (Høg et al. 2002). It is interesting to note at this point that several standards must be followed, specially those related to data packetisation. In order to fulfill this goal and to obtain an optimal solution for the timing and codification schemes, a set of models and software tools have been developed – which could also be applied for obtaining optimal solutions in other missions with similar operational principles and requirements. The amount of stellar objects per second to be observed by the instrumentation ranges from a very few up to some thousands, so an adaptive codification system is highly recommended. Furthermore, each of these measurements lasts for several seconds, so a pipelined operation is mandatory as well. A possible implementation of this structure is illustrated, as well as the several communication layers implied in the system.

The paper is organized as follows: section 2 gives an overview of the most relevant timing requirements for Gaia. Section 3 describes a set of generic designs for optimising the time tagging of the measurements and their transmission. Section 4 continues with these generic designs, offering now a set of models and software tools aimed to evaluate the occupation of a data link as a function of the several codification parameters. The optimal solution for the case of Gaia is described in section 5, which includes the telemetry structures to be generated, a possible implementation, and its predicted performance. This implementation could be used in any other instrumentation system with long measurement times and high rates – i.e., overlapped measurements. Finally, in section 6 we summarize the main results of our work and draw our conclusions.

## 2. Timing requirements of Gaia

The instrumentation of Gaia is based on CCD chips that will acquire the fluxes received from the stars and stellar objects. These will be the most important data to be transmitted, together with their time tags. These data, formed by sets of samples grouped as *patches*, will contain the image of the PSF of every object – usually emphasizing the resolution in the along-scan direction. The goal is to obtain their centroids with an extremely high accuracy and, hence, the precise position of every object. Although the photometric and spectrometric measurements will also be based on samples and star fluxes, we have focused our work on the astrometric instrument since it has the most stringent precision requirements.

Each of the two astrometric instruments will first detect the objects entering the focal plane using the astrometric sky mapper (ASM), which executes an image detection algorithm. Afterwards, the detected objects will be tracked all along the astrometric field (AF), obtaining the high-resolution measurements of their PSF. Since the geometry and behaviour of the focal plane will be known beforehand (and continuously calibrated during the mission), the knowledge of the detection coordinates will be enough for accurately linking the rest of measurements. Therefore, only the transit coordinates of the ASM will be required, besides the complete set of flux data. These transit coordinates, given by the Gaia detection system (GD), will be composed of both across-scan and along-scan components. The latter are the most important, since they will give the highest resolution, so we will focus on these.

While across-scan components contain the CCD and pixel of the ASM detected transit (with sub-pixel resolution), along-scan coordinates can be translated to time because of the deterministic, TDI-mode focal plane operation. These time data must be provided with equivalent sub-pixel accuracy, and transmitted in such a way that guarantees its unique reproduction on ground. Furthermore, the focal plane behaviour must also be reported to ground with the highest precision. For this, we must note at this point that the ultimate precision limit for Gaia is about 1µas or, equivalently, about 16ns of measurement time (Jordi et al. 2002). Time tagging should not increase unnecessary scattering, so a resolution better than 10ns is highly recommended. On the other hand, the detected ASM transit times must provide a resolution better than $1/200^{th}$ of the TDI period (736.6µs), in order to avoid quantisation errors on the GD output (which can offer a resolution of about $1/100^{th}$ TDI). Finally, all the time data fields must take into account the large time range envisaged for the mission. The possibility of a delayed launch, together with an extended mission, must be taken into account as well. Taking the recommended time reference 2010.0 (Bastian 2004), time data fields must be able to code up to 2025.

The transmission of such time data can eventually lead to an excessive telemetry channel occupation, since several hundredths of objects will be detected every second (and the focal plane behaviour should be continuously indicated). This implies that an optimised timing scheme must be developed. A possible solution is to transmit a reference time, say, once a second, and to transmit all of these time data as relative to this reference. This, on the other hand, could lead to an important data loss if some communication error occurs. Therefore, the timing and transmission scheme must be robust enough.

Finally, the acquisition and transmission of time data with the highest accuracy does not have any sense if these time data are not self-consistent. This implies that the payload itself must be built avoiding any inconsistencies and errors. A single spacecraft-based master clock is highly recommended for Gaia, which will probably be implemented with an Rb-type atomic clock (de Bruijne 2003). The distribution of this clock and its sub-products must be done using a symmetrical structure, implementing the transmission lines accurately. Bi-directional synchronisation signals may be exchanged if necessary, thus guaranteeing that all the instrumentation is operating with the same clock signal, fulfilling the phase difference requirements. Furthermore, the internal data fields must use the adequate resolutions: the use of better resolutions than the transmitted ones could also lead to inconsistencies and errors.

Clearly, the development of timing and transmission schemes must take into account all these requirements, thus leading to consistent schemes that, furthermore, imply a minimal occupation of the telemetry channel. The transmission scheme must also be prepared for applying lossless data compression afterwards.

## 3. Generic designs for optimised timing and transmission schemes

### 3.1   TIME DATA CODIFICATION METHOD

The case of Gaia is an example of instrumentation that generates a large amount of measurements, each requiring a very high resolution over a large time range. This implies that absolute time data fields would have prohibitive sizes. As illustrated in the top panel of Fig. 1, this can be considered an optimal solution for instrumentation generating just a few data blocks every second (which is not our case). A possible improvement is to group these data in *Data Sets*. These structures could be, for instance, of length 1 second, containing all the data *generated* (not *transmitted*) during that second. Then, a data set would contain a header with an absolute reference time, and the time tag of every data block would be relative to this reference. Thus, the required time range of these *relative* fields would drastically decrease, and so would do their sizes. This is illustrated in the central panel of Fig. 1, where it can be seen that the total amount of data decreases. Furthermore, the absolute time references transmitted every second could have a much lower resolution, thus indicating only the *coarse* time, while the relative time fields would be in charge of transmitting the *fine* time.

Yet another improvement can be applied to this timing scheme, as illustrated in the bottom panel of Fig. 1. The division of the data stream in *Data Sets* can be repeated, now dividing every data set in several *Time Slots*. This scheme is optimal only when large amounts of data blocks are generated every second (as it is our case). The basic principle is to transmit partial time references, which we name *time slot marks* (TSM), each relative to the last reference time. This procedure would reduce even more the effective time ranges, thus requiring even less bits for coding the time tags of every data block, relative to their corresponding TSM.

The time tags in Gaia are intrinsically quantised because of the TDI operation of the CCDs. More specifically, the image detection system reports these time tags based on the TDI charge transfer times of the focal plane. This makes possible to improve even more the timing and codification scheme, classifying any time data field in two main groups:

- Time data based on the *Satellite Time* (ST), this is, time fields that use second-based resolution. The sum of these fields offers a reference for the rest of time data. More specifically, they indicate the time at which the first sample of the current data set has been readout from the CCDs. We refer to this value as the *first effective readout finish time* (FEROFT). The reference time, with a resolution of 1 second and covering the full range of the mission, is the first contributor to this value. Nevertheless, the most critical contributor is the behaviour of the focal plane, which must be known with a resolution better than 10 ns.

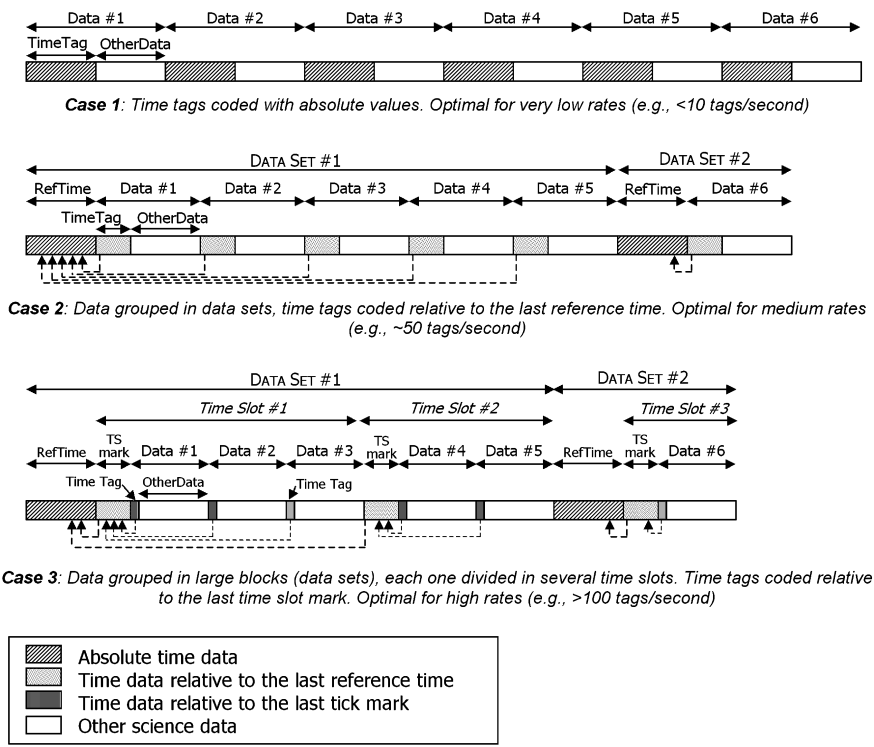

*Figure 1.*   Optimisation example for the time data codification, using time slots.

- Time data based on *TDI clock strokes*, this is, time fields that use a resolution of a fraction of the TDI period, and are reset at a given TDI clock stroke. The time slot marks (TSM) and the measurement time tags are included in this group. The latter are obtained from the image detection system, which reports the transit time of a given object over the ASM CCD. Hereof its name, *detected transit time* (DTT).

The TSM fields, transmitted at the beginning of every time slot, indicate the number of TDI clock strokes occurred since the *FEROFT*. At the same time, every DTT indicates the measurement time tag with respect to the last TSM, also as a number of TDI clock strokes. In this way, the time tags can be reconstructed with the best precision on ground, while requiring many less bits than in standard codifications. Furthermore, the beginning of a time slot will be coincident with its first DTT. Accordingly, the transmission of the latter can be safely suppressed. Therefore, the resolution of the TSMs will be the same as that of the DTTs, but their range will cover the whole data set – while the DTTs only have to cover the TSM length.

The maximum length of a time slot, which is related to the number of time slots to partition a data set in, will have to be determined in order to obtain optimal results. This maximum length will determine the code sizes of the DTT. The number of data blocks (astronomical objects, in the case of Gaia) that a time slot can contain will also have to be determined. It is important to note that there can be *holes* between time slots. That is, they do not necessarily have to continuously cover all the time line. Fig. 2 illustrates this final timing scheme recommended for Gaia, where the route to follow for decoding the actual time is indicated with dotted arrows. In this way, absolute times for every measurement can easily be obtained, just summing a few data fields. It is important to note at this point that this scheme is based only on the *detected* transit times: the total measurement time of each object is much longer than 1 second. This is why this system requires a pipelined implementation, as described below.

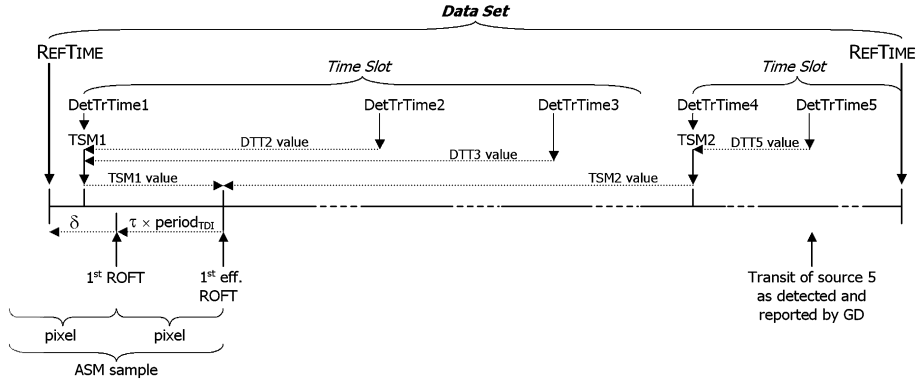

*Figure 2.* The time data codification method proposed for Gaia.

## 3.2  FULFILLING THE STANDARDS

It is clear that some compromises between the design of optimal schemes and the compatibility with existing systems must also be taken into account. That is, a minimum set of standards must be fulfilled, which include – at least – the telemetry standard. Also, it would be highly desirable to make possible an immediate transfer of most of the science data towards standard data exchange and data mining systems. It seems then obvious that time data is a good candidate for being standardised, and this is the goal of this section.

### 3.2.1  Time code formats

The Consultative Committee for Space Data Systems (CCSDS) recommends a set of time code formats (CCSDS 2002) for data interchange applications that are cross-supported between the agencies of the CCSDS. The timing scheme explained in section 3.1 includes the transmission of several time data fields, the most important of which is the reference time since it labels the primary data structures, this is, the data sets. Due to its low transmission rate (once a second) any standardisation overhead will be acceptable, so it can fulfill any required standard. We propose the CCSDS unsegmented time code (CUC), a counter of seconds since a given epoch. A Level-2 compliance is preferable (this is, agency-defined epoch), in order to make it compatible with the 2010.0 time reference. 4 octets of coarse time will give enough range (about 136 years). On the other hand, no octet of fine time will be required since we only need a resolution of 1 second. Hence, the total size of the code is 5 octets (including 1 octet of *preamble* field).

On the other hand, the $\delta$ and $\tau$ flags that form the FEROFT field (this is, the behaviour of the focal plane) will not fulfill any CCSDS standard. They will be transmitted at a higher rate, so any overhead introduced by standardisation should be avoided. Furthermore, $\tau$ is simply a Boolean value, so its standardisation would have no sense. On its hand, $\delta$ will be coded as a simple binary number, indicating the number of nanoseconds elapsed since the reference time.

While the reference time labels each data set, the several time slots will be labelled by the time slot marks (TSM). Therefore, their standardisation is recommended, in order to ease the exchange of such data structures with other applications. Our recommendation is a Level-3 compliance (agency-defined code), this is, a fixed number of octets conforming a binary number that represents the sub-second time within the data set. Furthermore, this number does increment monotonically with time (although not uniformly) and is reset only at the prefixed recycling instant. Its coding will be TDI-based, using a resolution of some hundredths of TDI period.

Finally, the detected transit times (DTT) should avoid any redundancy, since their transmission rate will be the highest one in the scheme. They will also be coded as TDI-based, but will not include any *preamble* field (required for standardisation), neither they will be restricted to an integer number of octets.

### 3.2.2 *Packet Telemetry*

The CCSDS has also developed a Packet Telemetry recommendation (CCSDS 2000), which will be used by the spacecraft telemetry formatter and, hence, must be fulfilled by this transmission scheme. The basic end-user data structure of this standard is the Source Packet (SP), which includes the identifier of the application process in charge of generating it. In our case, a possible and intuitive approach is to code every time slot as a source packet. We will assume this for the moment, so we can say that a source packet is the *coding equivalence* of a time slot – which contains data from several measurements. The several SPs are multiplexed into one or more virtual channels, which will finally be transmitted over the only physical channel available in our case. While the detailed contents of every SP to be generated will be described later, Fig. 3 illustrates the multiplexing process recommended in our case. As it can be seen, every instrument is considered as a separate Application Process (AP). The clock system in charge of generating the reference times is also considered an AP. Also the attitude and orbital control subsystem (AOCS), as well as the housekeeping subsystem, are transmitted as separate APs. These two data sources will be multiplexed in a separate virtual channel due to their different, spacecraft-based (instead of science-based) purpose.

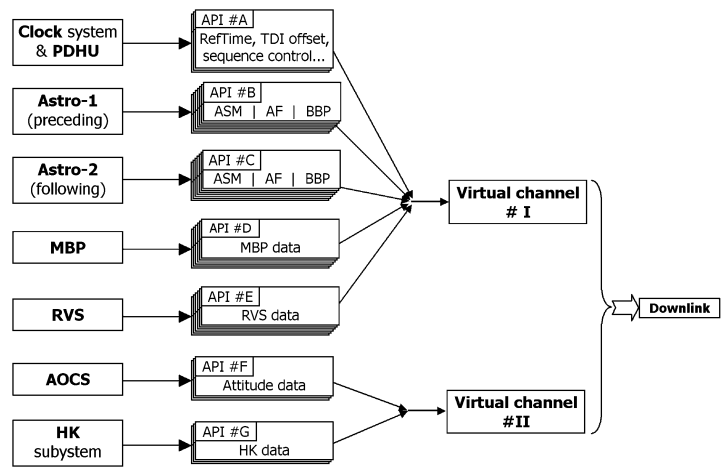

*Figure 3.*    The multiplexing process of the Gaia science telemetry.

Although there is only one Astro focal plane, data from measured stellar objects are separated depending on the telescope projecting them. On the other hand, Spectro (the other instrument) is broken down into a medium band photometer and a radial velocity spectrometer, which are implemented in two different focal planes. This multiplexing scheme fulfills the recommended time data codification method illustrated in Fig. 2, as well as the CCSDS recommendation of Packet Telemetry. Nevertheless, the payload data handling system may not generate the data sequentially, but may transmit some *low-* *priority* data at a later time. This could be done to ensure that all of the most relevant (*high-priority*) data is always transmitted, while the rest of data will only be transmitted if there is enough downlink capacity available. In this case, the timing structure of Fig. 2 will not be valid anymore, so an alternate structure should be developed for this *outdated* data. A possibility is to multiplex them in a separate virtual channel, using a similar structure based on data sets. Another possibility is to transmit absolute time data fields for every outdated measurement, since their transmission rate should be much lower – and, hence, their optimisation would not be so important.

Finally, and recalling one of the requirements in section 2, the coding and transmission scheme must be robust enough in order to avoid any error in the reception. Using a data set structure through a packet telemetry system may lead to errors in the reconstruction of time data fields. This can occur when the packet telemetry formatter changes the transmission order of the packets according to the telemetry channel availability, or it can be due to the process of data generation itself. In our case, all the measurements are based on a reference time transmitted in separate source packets (as illustrated in Fig. 3). Hence, a disordered packet may lead to a wrong time value reconstruction. In order to avoid this, we introduce a security system in our coding and transmission scheme, which is based on implicit *source sequence count* (SSC) values. They are simple sequence counters included in the header of every source packet, increasing independently for every application process. Therefore, if a data set header packet (including the reference time) indicates the next SSC values for every application process, a packet disordering can be safely solved. Fig. 4 illustrates this, where packets #6 and #7 are not transmitted in the expected order. The SSC values included in the data set headers solve this problem.

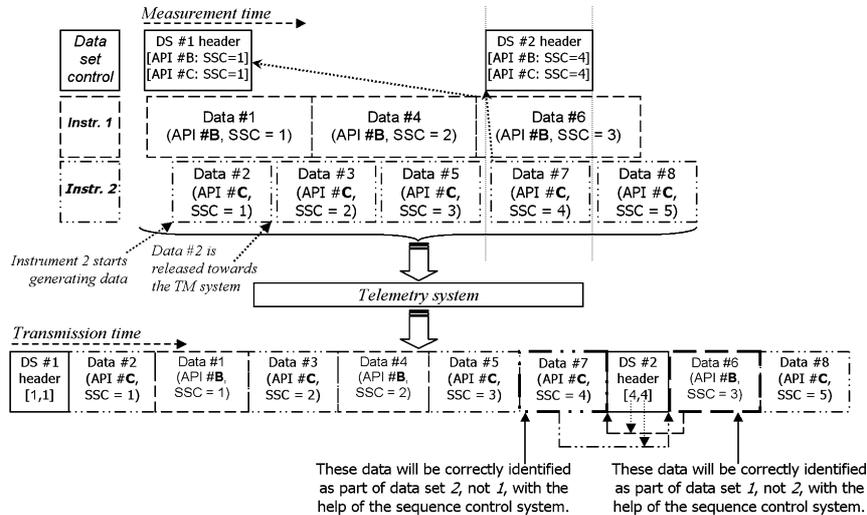

*Figure 4.*   Packet reordering process using the sequence control system.

We must also take into account that the loss of a source packet containing the data set header makes unusable all the data contained in that data set (since they will not have any reference time). This problem, as well as the problem of packet disordering, is a consequence of the use of such differential codifications. Nevertheless, we can also make the system robust enough to avoid these errors. The simplest solution is to include a redundancy, transmitting not only the SSC values of the current data set, but also those of the previous data set. In this way, since a reference time is generated every second (so its evolution is deterministic), the loss of a data set header will be solved with the extra data contained in the next data set header.

# 4. Modelling and simulation of data rates in a telemetry system

## 4.1 MODELLING

In order to obtain optimised codification and transmission schemes, the amount of telemetry data generated must be modelled as a function of the several parameters involved in this design. This must include not only the science data to be generated, but also the overheads introduced by the packetising process.

### 4.1.1 Codification parameters and mission parameters

We distinguish two main types of parameters: the codification parameters and the mission parameters. The latter cannot be freely modified, since they are determined by the mission requirements and definitions. On the other hand, codification parameters simply determine the way to encode science data, not altering its contents at all, so they can be modified at will. These are the parameters to optimise, trying to fulfill all the requirements listed in section 2 at the lowest cost in terms of telemetry consumption. Regarding the mission parameters, we will take in our case those values obtained from the latest design of the mission. It is important to note that, although we cannot modify the mission parameters, they can – and most times will – affect the final telemetry consumption.

In the case of Gaia we have introduced 6 different codification parameters, although some of them can be grouped or even handled as a single parameter. Flags $\delta$ and $\tau$, also referred to as *TDI Offset* and *ASM TDI Flag*, form the FEROFT field. If we want to ensure a perfect reconstruction of the behaviour of the several focal planes, such flags should be transmitted for each one of the CCDs. This would lead to an unnecessary downlink occupation, since the focal planes will be designed to operate as a whole. With this assumption in mind, transmitting only one FEROFT field per focal plane should be enough. Optimising it implies the determination of its optimal resolution, since its range will be determined by the TDI period of every focal plane. Therefore, and since the MBP and RVS focal planes (composing the Spectro instrument) use the same TDI period, only one optimisation for Spectro will be required – and another one for the Astro instrument. Nevertheless, we must remind that we focus on Astro, which is the most critical one – and which requires the most precise time measurements, at the level of a few nanoseconds.

The time slot marks (TSM) and detected transit times (DTT) are also codification parameters and must be optimised. By definition, both of them use the same TDI-based resolution, so they can be optimised as a single parameter. Furthermore, the first DTT of every time slot is not required, since its value is already indicated by the TSM – this is the clearest example of the similitude between the TSM and the DTT.

As indicated in section 3, the length of a time slot must be determined with simulations in order to obtain optimal results. This length determines the time range within a time slot, this is, the maximum value of a DTT field (since this field is coded relative to the last TSM). Therefore, the next codification parameter to optimise is referred to as the *maximum TSM offset* (MTO). Its value will determine the DTT range and, hence, the code size of the time data field that will be transmitted at the highest rate. This implies that the MTO will be a critical parameter in the simulations. Another interpretation of this parameter is the number of time slots that every data set will be partitioned in. More precisely, the MTO is proportional to the *inverse* of the number of time slots per data set.

Finally, the last codification parameter to optimise is the maximum number of measurements that can be stored in a single time slot. In our case it equals to the maximum number of *sources* (observed sources) that will be transmitted in a single time slot. Hence, we refer to it as the *maximum TS sources* (MTS). It will determine not only the range of a flag indicating the actual number of measurements in a source packet (this is, a time slot). A more important effect is that it will also determine the probability of packet loss (PPL), since larger packets have a higher probability of getting corrupted during their transmission down to the ground. It will be described thoroughly and evaluated later.

The most important of the mission parameters are the TDI periods (for Astro and Spectro), and the average star rate. The latter will be specially used as the *free* parameter with respect to which the several calculations will be optimised. Other mission parameters include the transfer frame size (indicated by the telemetry standard), the average downlink capability and the probability of frame loss (obtained by engineering teams), or the mission duration.

*4.1.2 Formulation*

For the sake of conciseness and clarity we will not go into the mathematical details of the formulation. Instead, the reader is referred to Portell et al. (2004), where the mathematical description of the model is discussed at length. However, we must remark at this point that the formulation developed for predicting the data rate from the instruments and the codification process is highly simplistic and averaged. This is, detailed models of the mission will not be included here, such as the nominal scanning law (NSL), the Galaxy model, or the on-board hardware limitations. Despite of this, the following formulation offers a good approximation to the telemetry occupation as it will be shown below. We also recall that we focus on the Astro instrument, although a few details of Spectro will also be treated.

The total data rate to be transmitted can be divided in three parts: fixed data rate, variable data rate from both Astro telescopes (to be most optimised), and equivalent data rate due to transmission errors:

$$Astro\_dr = dr\_fix + 2 \times dr\_var + dr\_err \qquad (1)$$

The fixed data rate (transmitted once a second) is determined by the data set headers, which include the reference time, the information on the behaviour of the focal plane and the packet sequence control:

$$dr\_fix = 48 + 50 + 8 \times \left\lceil \left( \left\lceil \log_2 \frac{TdiPerA}{TdiOffResA} \right\rceil + 2 \times \left\lceil \log_2 \frac{TdiPerS}{TdiOffResS} \right\rceil + 1 \right) \div 8 \right\rceil + 114 \qquad (2)$$

where *TdiOffResA* and *TdiOffResS* are the resolutions of the *TDI Offset* fields for Astro and Spectro, respectively. *TdiPerA* and *TdiPerS* are mission parameters, more specifically, the periods of the TDI clock for Astro and Spectro. On the other hand, the *variable* data rate is the sum of three main contributions: the source packet headers (including the TSM and actual number of measurements), the measurement time tags (this is, the DTTs), and the measurement data themselves:

$$dr\_var = \left( 48 + 8 \times \left\lceil \left( \left\lceil \log_2 \frac{DTTres}{TdiPerA} \right\rceil + 8 \right) \div 8 \right\rceil + \left\lceil \log_2 (MTS) \right\rceil \right) \times SP\_rate + \\ \left\lceil \log_2 (MTO \times DTTres) \right\rceil \times (StarRate - SP\_rate) + \frac{AvStarSize}{AvCmpRatio} \times StarRate \qquad (3)$$

where *MTO* is the maximum length of a time slot, *MTS* indicates the maximum sources that a time slot can contain, and *DTTres* is the resolution (in fractions of the TDI period) used to code the detected transit times (*DTT*). *AvStarSize* and *AvCmpRatio* are mission parameters, indicating the size of the measurement data themselves and the average compression ratio applied to them, respectively. Each of the three contributors of *dr_var* are transmitted at different rates, mainly depending on the *source packet rate* (*SP_rate*). Its value is determined by the source density of the field currently being observed, that is, by the so-called *star rate* (although not only stars are observed):

$$SP\_rate = \begin{cases} StarRate & (\text{for } StarRate \leq stdTSrate) \\ stdTSrate & (\text{for } stdTSrate < StarRate \leq stdTSrate \times MTS) \\ \left\lceil \dfrac{StarRate}{MTS} \right\rceil & (\text{for } StarRate > stdTSrate \times MTS) \end{cases} \quad (4)$$

where *StdTSrate* is the *standard time slot rate*, determined by the maximum TSM offset. Finally, the equivalent data rate due to transmission errors is given by the following expression:

$$dr\_err = AvDownlink \times PFL \times \frac{FS + SP\_size}{FS} \quad (5)$$

where PFL stands for the probability of frame loss (Pace 1998) and FS stands for the transfer frame size (ESA 1998). The average downlink capacity (*AvDownlink*) is also required, as well as an approximate value for the average source packet size (*SP_size*). For our purpose it is enough to calculate this as the average data rate divided by the source packet rate.

### 4.1.3  Model for a standard codification system

Although the model previously described offers a good approximation to the amount of data to be transmitted, it would be preferable to compare these results with standard codification systems. This is, absolute values are usually not enough for assuring that a system operates as expected, while a percentile of data rate saving (with respect to a standard encoding) is often much more intuitive and useful. In this section we show the formulation for obtaining the data rate generated by a *standard* measurement and codification system, equivalent to that one formulated in §4.1.2 but using a non-optimised encoding.

In the case of Gaia, a study for evaluating a prototype of database and data reduction system – Gaia data access and analysis study (GDAAS) – uses a standard codification, which only uses the data set optimisation. This will be our *standard* reference to which the optimised system will be compared. Its data rate model will be extremely similar, except for a few details – including a DTT coding with respect to the reference time, not to the time slot mark. The variable contribution is given by:

$$dr\_var_{EqGDAAS} = (48 + \lceil \log_2(MTS) \rceil) \times SP\_rate + \\ \left( \left\lceil \log_2 \frac{DTTres}{TdiPerA} \right\rceil + \frac{AvStarSize}{AvCmpRatio} \right) \times StarRate \quad (6)$$

where the source packets are generated following a simpler rule:

$$SP\_rate = \left\lceil \frac{StarRate}{MTS} \right\rceil \quad (7)$$

## 4.2 DEVELOPED SOFTWARE

Once the models were developed, we coded them as Matlab® *.m* files for being optimised, in order to occupy the minimum telemetry bandwidth. For this, a set of programs were developed, trying to offer user-friendly interfaces (mainly GUI-based, that is, graphical user interfaces). This software offers not only the possibility of changing the values of the several parameters – watching at their effect on the telemetry consumption results. It also offers coloured plots with automatic calculations of the optimal parameter curves. This is extremely useful with parameters like the MTO or the MTS, which are highly correlated with other parameters.

First of all, programs for an average-based parameter optimisation were developed. That is, fixing typical values in all of the mission parameters, and determining the best codification parameters under these *static* conditions. The left panel of Fig. 5 shows the main menu of *gaia_otdc_main.m*, the entry routine for this static optimisation. It contains edit boxes for manually changing the parameters, as well as numerical results after the computation of the telemetry models. It also displays several curves with the telemetry consumption as a function of the star rate (this is, the measurements per second), or with the telemetry savings compared to the standard codification. The buttons at the top left area load the optimisation routines, where sets of 2 parameters are optimised – as a function of the star rate. An example is shown in the right panel of Fig. 5, where the data rate is plotted with respect to the star rate and the maximum TSM offset. In this figure we can appreciate a white curve plotted over the three-dimensional plot, which marks the predicted optimal combination of parameter values. The user can modify all of the parameters at will, returning to the main menu, entering again in an optimisation routine for different parameters, and so on. At the end, the selected parameters and the resulting data rates (and savings) are numerically displayed.

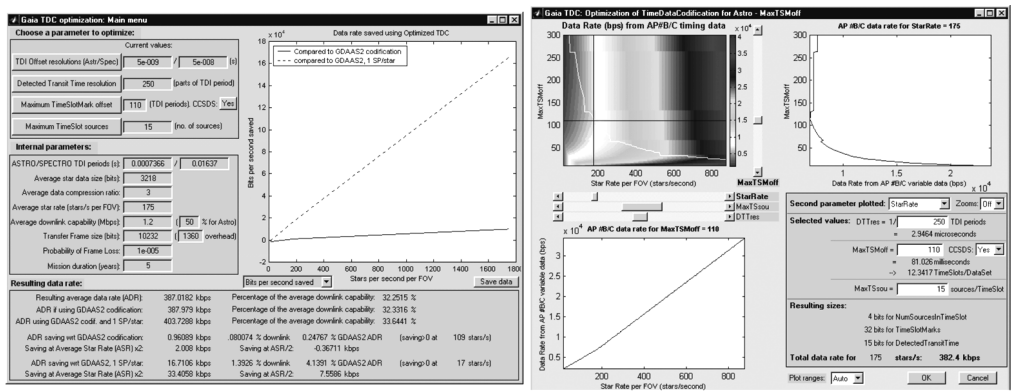

*Figure 5.*   Main menu and optimisation routine developed for selecting some codification parameters, using averaged modelling.

The interest of this software tool is not only related to Gaia – for which it gives excellent results. Its main interest is that telemetry models can easily be changed, since they are programmed in separate files. Therefore, this tool could be used for any other mission using similar codification principles. Even better, the timing and codification schemes can first be applied to a given mission, and afterwards this software tool can be run for obtaining the best results.

We must recall at this time that the operating conditions of Gaia will change dramatically during its operation. More specifically, Gaia will observe areas of the sky with large star density variations. This leads to very different star rates, ranging from a few stars per second up to 1500 stars per second (or even more). Nevertheless, the current design indicates that the average star density is about 174 stars per second. The same could happen to other missions performing systematic surveys, or simply operating under extreme conditions. If we use this static (averaged) value for obtaining a codification system, optimal results cannot be achieved. Actually, under certain conditions, this *optimised* system could offer even worse results than a standard system. This is the reason why we also developed another optimisation tool, shown in Fig. 6, operating with different parameters depending on the rate of the measurements. With this tool, we could convert this system into an *adaptive* codification system, trying to offer the best results under any operating condition.

Before developing this tool we had to decide which parameter should be adaptive. After a few tests with the average-based software, we verified that MTO is a crucial parameter. Its modification makes possible important telemetry savings, and its effect is highly dependant on the star rate. Therefore, it is clear that the time data codification (TDC) system must adapt the MTO depending on the density of the field being currently observed. We must recall at this point that scientific data is stored on-board before being transmitted to the ground. This makes possible an off-line analysis of the data, including a star density calculation, in order to obtain the best codification results.

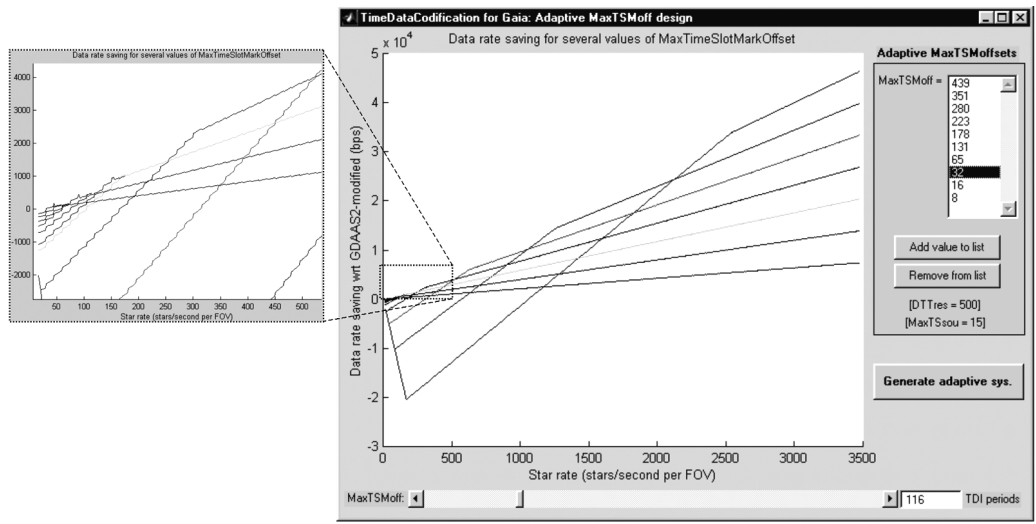

*Figure 6.*    Snapshot of the dialog for selecting an optimal adaptive codification system. A zoom of the low star rates area is included.

After entering *gaia_otdc_adap.m* under the Matlab® environment, the program first calculates a set of optimal values for MTO, stores them in a list, and plots the resulting telemetry saving. In other words, a telemetry saving curve is plotted for each of the MTO values added to the list. Furthermore, a *test curve* is also plotted using a user-selectable MTO value, which can be modified with a slider or an edit box. As it can be seen in Fig. 6, for each star rate range there is a given curve offering the highest telemetry saving. The user can test several values, adding them to the list or removing some of them, trying to obtain a minimum set of values offering the best results. When done, a list with the optimal MTO values – and their application ranges – is output by the program.

### 4.3    THE CASE OF GAIA: OPTIMAL PARAMETERS AND RESULTING PERFORMANCE

#### 4.3.1   Simulations of static codification

Using the average-based software with the adequate mission parameters and telemetry models for Gaia, always trying to obtain the best results with the lowest bandwidth penalty, we obtained the following codification parameters:

- Resolution of the TDI offsets for Astro: 5 ns (coded in 18 bits)
- Resolution of the TDI offsets for Spectro: 50 ns (coded in 19 bits).
- Resolution of detected transit times and time slot marks for Astro: $1/500^{th}$ of TDI period, this is, 1.4732μs. A time slot mark will be coded in 32 bits.
- Maximum sources in a time slot (or source packet): up to 15 sources. The actual number of sources will be indicated by a 4-bit flag.
- Maximum offset from the time slot mark: 116 TDI periods (85.4456ms), this is, a data set will contain about 12 time slots.

Using these static parameters we save more than 1 kbps with respect to a standard (yet slightly optimised) codification, being the main advantage that we are transmitting data with much higher time resolutions. Moreover the most important of them are coded with CCSDS time format standards. The source packet overhead is really small (about 0.3%), while we predict a total amount of data received with unrecoverable errors to be less than 0.005%.

#### 4.3.2   Simulations of adaptive codification

Adaptive coding software, as expected, produced better results than using static codifications. After a few tests, the set of MTO values indicated in Table 1 were obtained. Since only 7 values were required – so a 3-bit flag will be used for indicating the MTO value – the $8^{th}$ possibility was kept as reserved, in order to indicate an empty data set. It is important to note that the MTO value (and hence the codification performed) will be adapted once per second. This adaptation rate is enough for tracking the changes in the environmental conditions, while the penalty for the transmitted flag is negligible.

| Range (stars/s) | MTO flag value | Maximum TSM offset | Max TS length | Size of DTT fields |
|---|---|---|---|---|
| <61 | "000" | 8 TDIs | 386.0 ms | 12 bits |
| 61 to 120 | "001" | 16 TDIs | 193.0 ms | 13 bits |
| 120 to 256 | "010" | 32 TDIs | 96.5 ms | 14 bits |
| 256 to 526 | "011" | 65 TDIs | 47.9 ms | 15 bits |
| 526 to 1066 | "100" | 131 TDIs | 23.6 ms | 16 bits |
| 1066 to 2145 | "101" | 262 TDIs | 11.8 ms | 17 bits |
| >2145 | "110" | 524 TDIs | 5.9 ms | 18 bits |
| N/A | "111" | <void data set> | | |

*Table 1.* The several segments used by the adaptive codification system described in this work.

This adaptive software not only allows the calculation of the optimal system, but also the calculation of the *real* results when using a static system. In this case, using only one segment of MTO at 116 TDI periods, only 500 bps are actually saved. On the other hand, using the 7-segment adaptive system, more than 1.2 kbps are typically saved. The most interesting result is that the system can save telemetry even at very low star rates, such as only 31 stars per second – while the static system needed more than 100 stars per second in order to start saving telemetry. Also, this system can save up to 15 kbps or more when observing very dense star fields, which represents about 1% of the average downlink capability – only using a codification system, without any data compression yet.

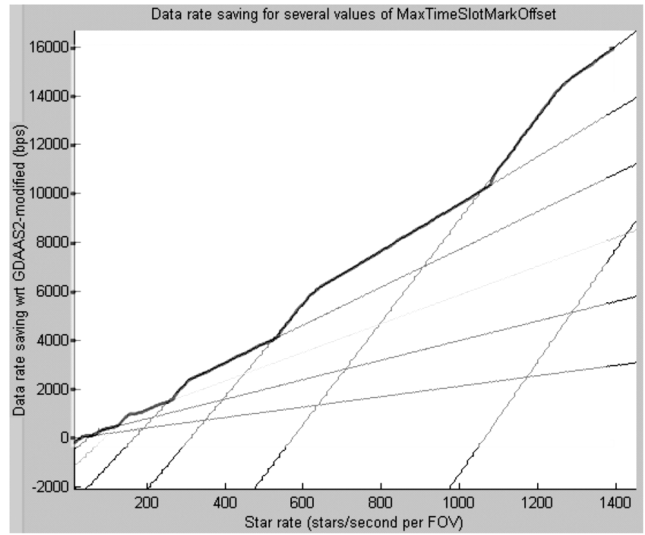

*Figure 7.*    Optimal data rate saving for any star rate, obtained with the adaptive codification system.

An equivalent impact on the scientific results is that there will be room for up to 10 additional measurements every second (or even more). Fig. 7 illustrates the performance of the adaptive coding system and the optimal data rate saving curve as a function of the star rate. The average overhead introduced by the source packets is about 0.3% again, but now it can be as low as 0.014%. The probability of packet loss is typically lower than $3\times10^{-5}$, which is equivalent to a total amount of unrecoverable data of about 0.003% of the total data transmitted. The predicted data rate, in average, is about 1.18 Mbps without compression – 398 kbps if we assume a typical compression ratio of 3 (Portell et al. 2002).

## 5. Implementation and integration in communication systems

### 5.1    SOURCE PACKETS TO BE GENERATED. THE CASE OF GAIA

With the results obtained using the simulations, the source packets to be generated by our data handling system are finally defined and fixed. These packets will be ready to be fed into the standard Packet Telemetry system for being transmitted, not only in CCSDS-compliant systems but also in systems based on the ESA Packet Telemetry Standard (ESA 1998). This is possible because the few modifications introduced by this variation do not need to be applied to our data structures.

Two main types of packets have been defined in our work, the first one of which is the data set header – used both for the Astro and the Spectro instruments. The contents of this packet must include, first of all, the reference time with a resolution of 1 second and the format previously recommended (§3.2.1). Data indicating the current behaviour of the focal plane must also be included here, as well as the *MTO* flag – indicating the current mode of the adaptive codification system. The sequence control system, reporting the SSC values with which the current and previous data set contents start, are included afterwards. Finally, although not considered in this work, any other ancillary data could be included at the end. It may include a complete report of the focal plane behaviour (for each of its CCDs), variations in the actual TDI period, etc. The rest of contents of the packet are determined by the standard or assigned by our recommendation, such as the application process identifier. We must remark that a packet must contain an integer number of octets, so *dummy* bits must be added for filling the *holes*. Fig. 8 shows the source packet contents and sizes, where it can be seen the shade indicating a dummy bit. We can also see that the MTO flag is transmitted both for the current and the previous data set, in order to ensure its correct reception – just as with the sequence control system. With a transmission only once a second, these packets occupy only 264 bits per second, despite of those redundancies.

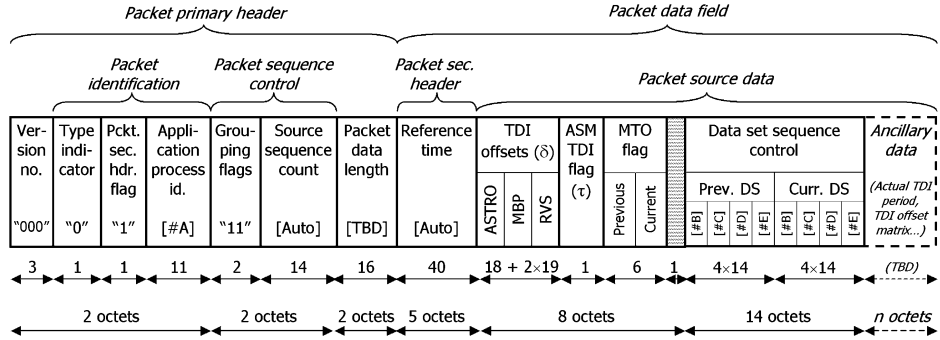

*Figure 8.*   Packet structure and contents for AP #A (data set header).

The other packet type defined in our work is only related to the Astro instrument, on which we have focused our efforts. As it can be seen in Fig. 9, the first field introduced by our design is the time slot mark (TSM), with the CCSDS-compliant format specified in section 3.2.1. Next to it, the actual number of sources transmitted in the current packet is indicated. It must be noted that we must subtract 1 to it in order to obtain the number of measurement time tags (this is, DTT fields) transmitted afterwards, since the first DTT is already indicated by the TSM. Finally, the measurement data (specially flux data) are transmitted – including the adequate compression system.

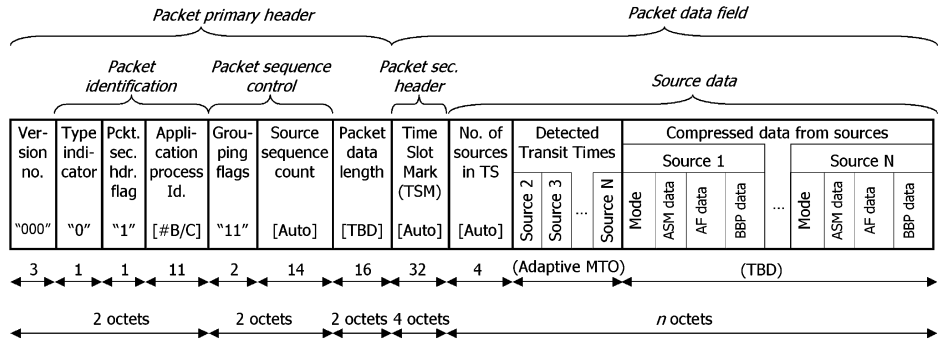

*Figure 9.*   Packet structure and contents for AP #B/C (Astro data).

It is worth pointing out at this point that this is not the usual operation in data handling and transmission systems: the data is usually compressed *before* packetising it. This is, we use an *inverse packetising*. Our recommendation operates in this way because, first of all, it eases and emphasizes the previous optimisation process. We also separate in this way some of the measurement data (this is, the timing data) for optimising it as explained in this paper. The last reason, but surely the most important one, is that the resulting measurement data stream gets more uniform. In the case of Gaia, specifically, the remaining data is mainly flux data. Preliminary studies (Portell et al. 2001; Portell et al. 2002) demonstrate how compression ratios significantly increase when applied to data with similar natures and formats, as it could be expected. Therefore, this codification process not only saves a significant amount of telemetry by itself, but also makes possible a much more important saving when applying data compression afterwards.

5.2    IMPLEMENTATION OF THE DATA COLLECTION AND CODIFICATION PROCESSES

We have defined a set of designs and procedures for time tagging measurements, coding them and packetising them in an optimised way. At a first sight, when only the *generation* time of the measurements is taken into account, it may seem simple to implement. This time value is used for labelling each observation, and is the only time data used during all the process. Other time data is also used – for reporting the behaviour of the instruments – but, at the end, this value is used for obtaining more accurate time tags of the measurements. The problem appears when it is realized that each of the transits of the astronomical objects over a focal plane lasts several seconds (57 seconds in the case of Astro). Therefore, this is the time after which the complete measurement data block will be released. This, combined with the large amount of observations performed every second, complicates the data collection process. The best solution for this is a pipelined implementation, based on several processes and data buffers operating simultaneously – each with a portion of the measurements currently being done. An overview of this is illustrated in Fig. 10.

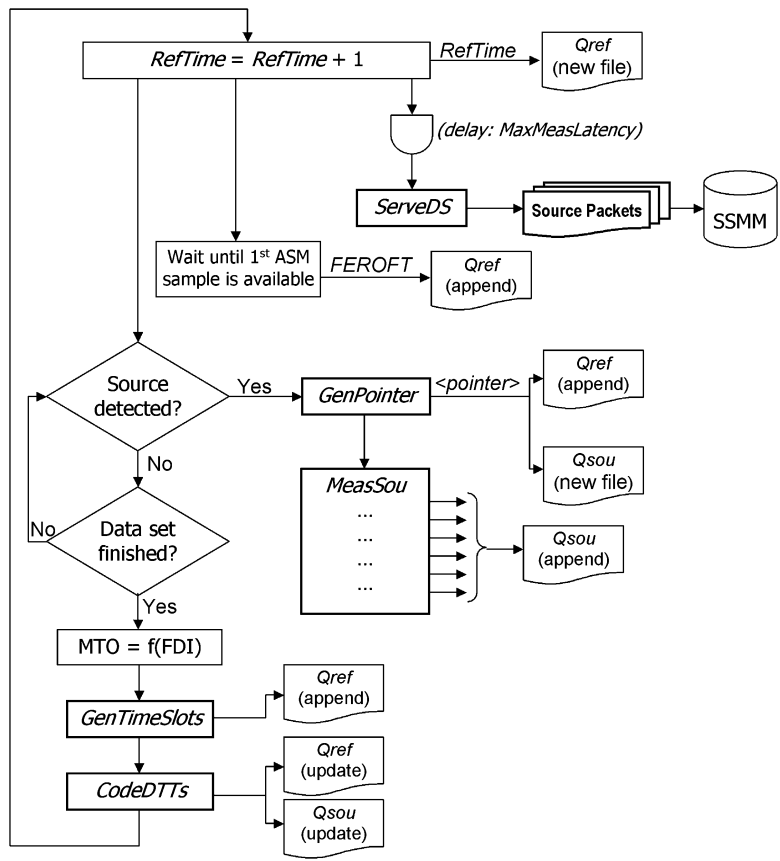

*Figure 10.*   Flux diagram of an adaptive telemetry coder.

Two main data buffers are used during this pipelined processing, both of them operating as FIFO queues. One of them, $Q_{ref}$, is used for referencing the several data currently being measured. A reference time will be entered there every second for marking the beginning of a new data set, as well as the *FEROFT* flags of the focal plane. Afterwards, a pointer to each of the several sky objects detected (or, generically, the several new measurements) will be generated and stored in this queue. This pointer may be as simple as the corresponding detected transit time (in full resolution and absolute mode). This time will also be stored in the second data buffer, $Q_{sou}$, which will store the several measurement data themselves. In the case of Gaia, the process in charge of collecting and storing these data is called *MeasSou* (measure sources), as seen in Fig. 10.

Calculations for the adaptive system will be performed at the end of the data set. A flag indicating the current measurement rate will be calculated, which we name *field density index* (FDI) since Gaia will observe whole stellar fields. From this, the corresponding MTO flag will be calculated, and the adequate time slots will be generated. They will be stored in the reference queue ($Q_{ref}$), together with the corrected DTT values – which will also be updated to the sources queue ($Q_{sou}$). With this, the adequate time data codification is already obtained, although the measurements have just begun.

Since our full set of instrumentation contains elements releasing measurement data at different latencies, all of them must be unified for being correctly fed into the telemetry stream. This is, for keeping the adequate sequence marked by the data sets and time slots. In our case (Jordi et al. 2003), we have the Astro instrument releasing data 57 seconds after its first generation, and the Spectro instrument 89 seconds after (MBP) or 145 seconds after (RVS). The best solution is to take the longest latency time (this is, 145 seconds), which we name *MaxMeasLatency* in the figure, and retain all of the measurements during this time. Afterwards, the data set can be *served* sequentially, taking all the data indicated by the reference queue and generating the corresponding source packets. All these data can finally be stored in the solid state mass memory (SSMM), waiting for the next contact with the ground station for transmission.

### 5.3    COMMUNICATION LAYERS

The several recommendations on Packet Telemetry (CCSDS 2002; ESA 1998) already illustrate the end-to-end operation of such communication systems, indicating the data structures implied. It includes the source packets, the telemetry packets (equivalent to the source packets in our case) and the transfer frames. Here we have designed the way the data handling system must fill two main types of source packets. Nevertheless, a global view of our complete data and communications system has also been elaborated, the result of which is shown in Fig. 11. This scheme illustrates the several communication layers implied, from the instrumentation to the transmitter – and their corresponding layers at reception. We must note at this point that several simulators are being developed in order to design and test the several communication layers, as well as the data generation and compilation.

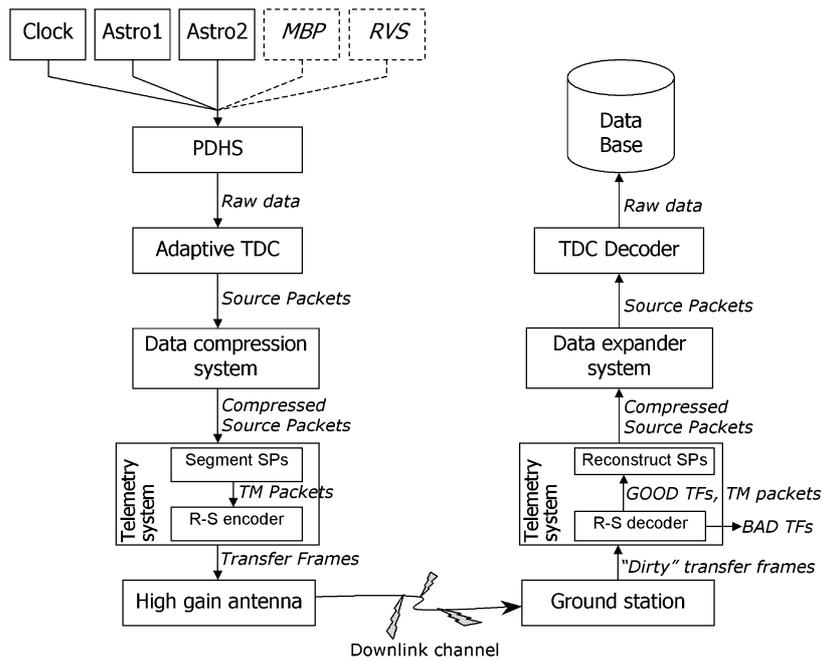

*Figure 11.*   Data communication layers in Gaia.

## 6. Conclusions

The main features of the Gaia mission have been presented, focusing on its payload and instrumentation, and more specifically on its astrometric instrument. An optimised time data codification method has been presented, fulfilling strict timing requirements and resulting in a reduced amount of data generated. This issue is crucial in instrumentation systems like Gaia, which performs a large amount of high resolution measurements but has a limited downlink capacity. The fact that the measurements are tagged with implicitly discrete time values has helped in obtaining a lower amount of raw data to be transmitted. This timing scheme obviously required a corresponding codification scheme. In our work we have described a CCSDS-compliant solution, capable of offering an excellent scientific return at the lowest price in terms of telemetry occupation. A few codification parameters had to be selected in order to achieve this. Specific software tools have been developed, which are modular enough for modifying the telemetry models and, thus, being applicable to other particular cases. The simulations performed with these tools led to schemes that give the best results under any conditions, thanks to the adaptive operation selected for our codification system. Furthermore, the transmission scheme is robust enough in front of unrecoverable errors in the reception stage. One of the interesting details of this recommended scheme is that measurement data will be packetised before compressing it. This inverse solution not only offers better optimised schemes, but also makes possible the compression of measurement data at higher ratios, since data of the same kind is previously packetised together. It also makes possible eventual deactivations of the data compression system in case of errors, offering at the same time lower telemetry occupations than with standard systems. Finally, a recommended implementation has been described. This has to operate as a pipeline, since several high latency measurements are performed every second – so all of them overlap in time. The communication layers implied in all these systems are also shown, from the instrumentation to the final database. Summarizing, the use of these designs leads to a better scientific return with the same communications channel, only requiring a few modifications in the data handling systems. Although we have focused our work on the Gaia mission, most of these systems can easily be applied or adapted to other missions with similar operational principles.